%% file: aph.tex
\documentclass[iop]{emulateapj-rtx4}
\usepackage{natbib}
\usepackage{comment}
\usepackage{lscape}
\usepackage{amsmath}
\usepackage{hyperref,breakurl}

\shorttitle{THE ROLE OF MERGERS ON GALAXY RADIO SPECTRA }
\shortauthors{MURPHY}
\slugcomment{Submitted to ApJ June 7, 2013; Accepted August 15, 2013}
\begin{document}
%\title{On the Steep, High-Frequency Radio Spectra Local Infrared-Bright Starburst Galaxies: The Role of Merger Stage on Galaxy Radio Spectra}
\title{The Role of Merger Stage on Galaxy Radio Spectra in Local Infrared-Bright Starburst Galaxies}

\author{Eric~J.~Murphy}
\affil{Infrared Processing and Analysis Center, California Institute of Technology, MC 220-6, Pasadena CA, 91125, USA; emurphy@ipac.caltech.edu\\
Observatories of the Carnegie Institution for Science, 813 Santa Barbara Street, Pasadena, CA 91101, USA}
%\affil{Observatories of the Carnegie Institution for Science, 813 Santa Barbara Street, Pasadena, CA 91101, USA}

\begin{abstract}  
An investigation of the steep, high-frequency (i.e., $\nu \sim 12$\,GHz) radio spectra among a sample of 31 local infrared-bright starburst galaxies is carried out in light of their {\it HST}-based merger classifications.  
Radio data covering as many as 10 individual bands allows for spectral indices to be measured over three frequency bins between $0.15-32.5$\,GHz.    
Sources having the flattest spectral indices measured at $\sim$2 and 4\,GHz, arising from large free-free optical depths among the densest starbursts, appear to be in ongoing through post-stage mergers.  
The spectral indices measured at higher frequencies (i.e., $\sim$12\,GHz) are steepest for sources associated with ongoing mergers in which their nuclei are distinct, but either share a common stellar envelope and/or exhibit tidal tails. 
These results hold after excluding potential AGN based on their low 6.2\,$\mu$m PAH EQWs.  
Consequently, the low-, \mbox{mid-}, and high-frequency spectral indices each appear to be sensitive to the exact merger stage. 
It is additionally shown that ongoing mergers, whose progenitors are still separated and share a common envelope and/or exhibit tidal tails, also exhibit excess radio emission relative to what is expected given the far-infrared/radio correlation, suggesting that there may be a significant amount of radio emission that is not associated with ongoing star formation.   
The combination of these observations, along with high-resolution radio morphologies, leads to a picture in which the steep high-frequency radio spectral indices and excess radio emission arises from radio continuum bridges and tidal tails that are not associated with star formation, similar to what is observed for so-called ``taffy" galaxies.  
This scenario may also explain the seemingly low far-infrared/radio ratios measured for many high-$z$ submillimeter galaxies, a number of which are merger-driven starbursts.  

\end{abstract}

\input{tbl-1}

\section{Introduction}
Encoded in the radio spectra of star-forming galaxies, which are typically well characterized by a power law ($S_{\nu} \propto \nu^{-\alpha}$), lies information on the thermal and non-thermal energetic processes powering them.  
Both thermal and non-thermal emission processes are typically associated with massive star formation, underlying the basis for the well known far-infrared/radio correlation \citep{de85, gxh85, jc92}.  
Far-infrared emission arises from re-radiated UV/optical photons that heat dust grains surrounding massive star-forming regions.
The young, massive O/B stars in such regions, whose lifetimes are $\la 10$\,Myr, produce ionizing radiation that is proportional to the amount of free-free emission.  
Stars more massive than $\ga 8\,M_{\sun}$ end their lives as supernovae (SNe), whose remnants (SNRs) are thought to be the primary accelerators of cosmic-ray (CR) electrons, which emit synchrotron emission as they propagate through a galaxy's magnetized interstellar medium (ISM).  

%Galaxy radio spectra are typically well characterized by a power law, where $S_{\nu} \propto nu^{-\alpha}$.  
The non-thermal emission typically dominates the free-free emission at frequencies $\la 30$\,GHz \citep{jc92}, having a relatively steep spectrum \citep[i.e., $\alpha \approx 0.83;$][]{nkw97}.    
Thermal bremsstrahlung (free-free) emission, on the other hand, has a much flatter spectrum ($\alpha \approx 0.1$), making it difficult to separate this component from the non-thermal emission.  
Thus, at frequencies $\ga 30$\,GHz, where the thermal fraction starts to become large, radio observations should become robust measures for the ongoing star formation rate in galaxies \citep{ejm11b,ejm12b}.  
However, this has been shown to not necessarily be the case for a number of local luminous infrared galaxies (LIRGs), whose infrared (IR; $8-1000\,\mu$m) luminosities exceed $L_{\rm IR} \ga 10^{11}\,L_{\sun}$.  

In a number of these infrared-bright starbursts, their high-frequency (i.e., $\ga 10\,$GHz) radio spectra are much steeper than expected for an increased thermal fraction, and in some cases, even show possible evidence for spectral steepening \citep{msc08, msc10, akl11b}.  
Understanding the physical underpinnings driving this behavior can greatly help with the interpretation of radio observations for higher redshift starbursts, which is important given that   infrared-luminous galaxies appear to be much more common in the early universe and dominate the star formation rate density in the redshift range spanning $1 \la z \la 3$, being an order of magnitude larger than today \citep[e.g.,][]{ce01,el05,kc07,ejm11a,bm13}.

\input{tbl-2}

In the local universe, it is well-known that LIRGs and ultraluminous LIRGs (ULIRGs; $L_{\rm IR} \ga 10^{12}\,L_{\sun}$) appear to be undergoing an intense starburst phase.  % of star formation.  
Within these systems are compact star-forming regions that have been been triggered predominantly through major mergers 
\citep[see e.g.,][]{lee87,lee88,lee89,lee90,dbs88a,dbs88b,twm96,sv95,sv97,sv02}.  
Major mergers have the ability to significantly complicate the interpretation of observed radio properties when individual systems are not resolved, the classic case being the so-called ``taffy" galaxies \citep{jc93,jc02}.  
When unresolved, the systems appear to have nearly a factor of $\sim$2 more radio continuum emission relative to what is expected giving the far-infrared/radio correlation, as well as unusually steep (i.e., $\alpha \approx 1.0$) radio spectra.  
However, when resolved, it is clearly found that the integrated radio properties are driven by radio continuum emission that forms a bridge connecting the galaxy pairs; 
the radio continuum emission in the bridges are characterized by a steep spectrum and contain roughly an equal amount of emission as the individual galaxies, which have radio spectral indices and far-infrared/radio ratios typical of normal spiral galaxies \citep{jc93,jc02}.  
%While the ``taffy"-like systems are seemingly rare, 

In this paper, an explanation is presented for occurrences of steep radio spectra observed in local ULIRGs within the context of there merger stage.  
The paper is organized as follows. 
In \S2 the sample and data used in the analysis are presented.  
In \S3 the major results are described, and then discussed in \S4 to piece together a self consistent picture describing the radio properties for this sample of local starbursts.  
Finally, the main conclusions are summarized in \S5.

\section{Data and Analysis}
\label{sec-data}
The galaxy sample being analyzed here is drawn from sources in 
%The galaxies in this sample are drawn from 
the {\it IRAS} revised Bright Galaxies Sample \citep{bgs89,rbgs03} having 60\,$\mu$m flux densities  larger than 5.24\,Jy and far-infrared (FIR; $42.5 - 122.5\,\mu$m) luminosities $\geq 10^{11.25}\,L_{\sun}$.  
The 40 systems that meet these criteria were originally imaged by \citet{jc91} at 8.4\,GHz with 0\farcs25 resolution.   
For 31 of these sources, \citet{msc08} presented additional new very large array (VLA) 22.5\,GHz data (obtained in D-configuration), along with new 8.4\,GHz data for 7 objects (obtained in C-configuration) and archival observations to increase the radio spectral coverage of each system.  
Additional lower frequency data at 244 and 610\,MHz were obtained and presented in \citet{msc10}, and for 14 and 13 sources existing 6.0 and 32.5\,GHz measurements, respectively  from \citet{akl11b} are used.   
The present analysis focuses on these 31 galaxies (see Table \ref{tbl-1}) having well sampled radio spectra between 0.15 and 32.5\,GHz (typically 6 bands with as many as 10 including some combination of data observed at 0.151, 0.244, 0.365, 0.610, 1.4, 4.8, 6.0, 8.4, 15.0, 22.5, and 32.5\,GHz).   
While these radio data span over an order of magnitude in frequency and have varying spatial resolutions, missing short-spacing data will not significantly affect the results given that the highest frequency radio data should be sensitive to  $\gtrsim$1\arcmin~angular scales, much larger than the typical emitting regions for this sample of infrared-bright galaxies.  

For NGC\,3690, UGC\,08387, UGC\,08696, and IRAS\,F15163+4255, the 22.5\,GHz flux densities of \citet{msc08} are significantly lower than the 32.5\,GHz flux densities presented in \citet{akl11b}.  
These discrepancies may be associated with the difficulties of measuring flux densities through the pressure-broadened 22.3\,GHz line of atmospheric water vapor, and therefore the 22.5\,GHz flux densities for these 4 sources are not included in the present analysis.  
Additionally, the 15\,GHz flux density UGC\,08387, which was reduced from archival VLA data by \citet{msc08}, is found to be discrepant with the combination of lower frequency radio data and the 32.5\,GHz flux density of \cite{akl11b}, and is therefore not used.  
For IRAS\,F15163+4255, the 4.8\,GHz flux density measured by \citet{gc91} using the 91\,m telescope in Green Bank is found to be significantly discrepant with the 6.0\,GHz flux density of \citet{akl11b}, and is dropped from the analysis as well.    
% Additionally, the, 15Thus, 
%{\bf say something about not using 4.9 and 22.5GHz data for VV705 and other galaxies since significantly discrepant with Leroy data}.  

{\it IRAS}-based far-infrared fluxes and logarithmic far-infrared-to-radio ratios at 1.4 and 8.4\,GHz, each of which were reported in \citet{msc08}, are also used in the analysis.    
The far-infrared flux, $F_{\rm FIR}$, is estimated using {\it IRAS} 60 and 100\,$\mu$m flux densities and the relation given by \citet{gxh85} such that, 
\begin{equation}
\left(\frac{F_{\rm FIR}}{\rm W\,m^{-2}}\right) = 1.26\times10^{-14}\left[\frac{2.58 f_{\nu}(60\,\micron) + f_{\nu}(100\,\micron)}{\rm Jy}\right].  
\end{equation}
Logarithmic $F_{\rm FIR}$/radio ratios at frequency $\nu$ are also defined using the relation given by \cite{gxh85} such that 
\begin{equation}
q_{\nu} = \log\left(\frac{F_{\rm FIR}}{3.75\times10^{12}~{\rm W\,m^{-2}}}\right) - \log\left(\frac{S_{\nu}}{\rm W\,m^{-2}\,Hz^{-1}}\right). 
\end{equation}
Each of these values is given in Table \ref{tbl-1} for each source.  

%The parent sample for this investigation is drawn from the results of \citet{jc91}, who mapped the 40 galaxies included in the {\it IRAS} revised Bright Galaxies Sample \citep{bgs89,rbgs03} having far-infrared (FIR; $42.5 - 122.5\,\mu$m) luminosities $\geq 10^{11.25}\,L_{\sun}$ at 8.44\,GHz with 0\farcs25 resolution.  
%Of the resolved galaxies, radio spectral indices measured between 1.49 and 8.44\,GHz are available for 36 sources \citep{jc91}, which defines the sample used in the present analysis.  
%Additional components were resolved by the 8.44\,GHz data, for which we assume the same spectral index when comparing with their mid-infrared properties.  
%We additionally make use of the derived 8.44\,GHz sizes and brightness temperatures available for 28 of the sources, which are all given in Table \ref{tbl-1}.  
%As shown by \citet{jc91}, sources having 8.44\,GHz brightness temperatures $\ga10^{4.5}$\,K cannot be sustained by star formation alone, and must be powered by nuclear ``monsters" (AGN).  

The mid-infrared spectral properties were collected by the {\it Spitzer} Infrared Spectrograph \citep[IRS;][]{jh04} as part of the Great Observatories All-Sky LIRG Survey \citep[GOALS][]{lee09}, 
and are taken from \citet{ss13a}.  %Stierwalt et al. (2012) and S. Stierwalt et al. (2012, in preparation).  %\citep{ss13}.  
For 2 sources, IRS observations missed the nucleus of the galaxy (IRAS\,F01417+1651, IRAS\,F03359+1523).  
%The total IR ($8-1000\,\mu$m) luminosities are taken from \citet{lee09}.  
%However, for sources having multiple components as resolved by the IRS, the fractional IR luminosity from each component is determined by the fraction of resolved 24\,$\mu$m emission.  
In the analysis 6.2$\mu$m polycyclic aromatic hydrocarbon (PAH) equivalent widths (EQWs) and 9.7$\mu$m silicate strengths \citep[see][]{ss13a} are used.  %as defined by \citet{hs07}.  
Similar to \citet{ejm13}, the measured 6.2\,$\mu$m PAH EQWs is used as an active galactic nuclei (AGN) discriminant.    
%The 6.2\,$\mu$m PAH EQW has been used to indicate 
Sources hosting AGN typically have very small PAH EQWs \citep[e.g.,][]{rg98,lee07}.
Specifically, starburst dominated systems appear to have 6.2\,$\mu$m PAH EQWs that are $\ga0.54\,\mu$m \citep{bb06}, while AGN have 6.2\,$\mu$m PAH EQWs are $\la 0.27\,\mu$m.  
In this paper it is assumed that all sources having PAH EQWs $\geq 0.27\,\mu$m are primarily powered by star formation.  
Given that the PAH EQW is used as an AGN discriminant, we exclude both sources missing IRS data in the present analysis when focusing on star formation dominated systems.  
%, leaving a total of 29 galaxies.  
The silicate strength at 9.7\,$\mu$m is defined as $s_{9.7\micron} =  \log(f_{9.7\micron}/C_{9.7\micron})$, where $f_{9.7\micron}$ is the measured flux density at the central wavelength of the absorption feature and $C_{9.7\micron}$ is the expected continuum level in the absence of the absorption feature.  
For two sources (UGC\,04881 and NGC\,3690), IR luminosity-weighted averages of individual IRS measurements from each nuclei are used.  
%where the level of the continuum would be in the absence of the absorption feature.  
%, based on an extrapolation to the surrounding continuum.  
%We additionally adopt the IR luminosities given by \citet{jh10}.  
%While the 0\farcs25 radio maps are at a resolution that is significantly higher than the mid-infrared IRS data (i.e., 3\farcs6), this should not affect our analysis or conclusions given that the central sources seem to dominate the luminosities of each system.  
%Furthermore, the radio data themselves should be sensitive to extended emission on scales up to $\approx$5\arcsec, which is larger than the resolution of the IRS data.  

This analysis additionally makes use of merger classifications based on available {\it Hubble Space Telescope} ({\it HST}) imaging for 29 galaxies taken as part of an {\it HST}/ACS survey of the GOALS sample \citep[][A.S. Evans et al. 2013, in preparation]{dck13}.  
The merger classifications were assigned using both {\it HST} optical [i.e., Advance Camera for Surveys (ACS) $B$- (F435W) and $I$-band (F814W)] and near-infrared [i.e., Near Infrared Camera and Multi-Object Spectrometer (NICMOS) $H$-band (F160W)] imaging.  
The individual mergers stages are classified by an integer value ranging from 0 to 6, and are described in detail by \citet{haan11}. 
Briefly, the numerical classifications are defined in the following way:  0 $=$ non-merger, 1 $=$ pre-merger, 2 $=$ ongoing merger with separable progenitor galaxies, 3 $=$ ongoing merger with progenitors sharing a common envelope, 4 $=$ ongoing merger with double nuclei plus tidal tail, 5 $=$ post-merger with single nucleus plus prominent tail, and 6 $=$ post-merger with single nucleus and a disturbed morphology.  
For two sources, merger classifications are not available.  
{\it HST} data were not taken for NGC\,6286, and the {\it HST}/ACS data saturated for the observations of UGC\,08058 (Mrk\,231).  
Each of these properties is given in Table \ref{tbl-1}.  

\begin{comment}
Stellar masses are taken from \citet{vu12}, who derived estimates by fitting optical through near-infrared spectral energy distributions, as well as assuming a fixed mass-to-light ratio and $H$-band flux densities, for both Salpeter \citep{salp55} and Chabrier \citep{chab03} IMFs.  
The masses adopted here are the mean of these estimates after first converting them to a Kroupa \citep{pk01} IMF using the following statistical relations as given by \citet{mbolz10}: \(\log{M_{*}({\rm Chab})} \approx \log{M_{*}({\rm Salp})} - 0.23\) and \(\log{M_{*}({\rm Krou})} \approx \log{M_{*}({\rm Chab})} + 0.04\).  
Each of these properties is given in Table \ref{tbl-1}.  

Infrared luminosities are converted into a star formation rate using the calibration of \citet{ejm11b}, which assumes a Kroupa IMF, such that,
\begin{equation}
\left(\frac{\rm SFR}{M_{\sun}\,{\rm yr^{-1}}}\right) = 1.48\times10^{-10}\left(\frac{L_{\rm IR}}{L_{\sun}}\right).
\end{equation}
We note that the coefficient here is 1.23 times larger, but within the scatter, of the empirically derived relation between IR and radio free-free star formation rates \citep{ejm12b}.  
\end{comment}

\begin{figure}
\epsscale{1.1}
\plotone{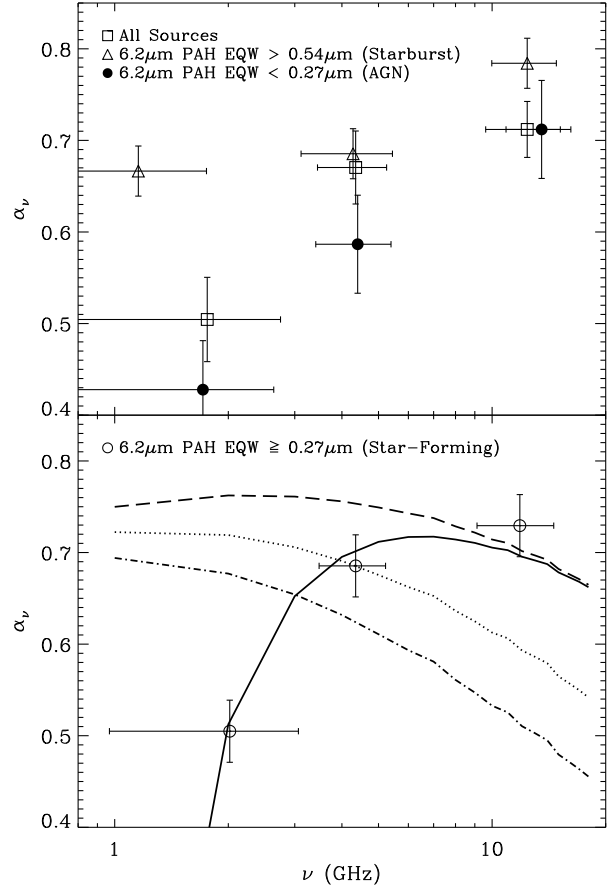}
\caption{The median radio spectral indices calculated at three different frequency bins listed in Table \ref{tbl-2}.  
The horizontal error bar illustrate the standard deviation among the median frequencies per bin, while the vertical error bars illustrate the error on the median spectral index.  
In the top panel, sources are binned based on their 6.2\,$\mu$m PAH EQWs, where starbursts (i.e., 6.2\,$\mu$m PAH EQW $>$ 0.54\,$\mu$m) and AGN (i.e., 6.2\,$\mu$m PAH EQW $<$ 0.27\,$\mu$m) show statistically different average radio spectral indices in the low- and mid-frequency bins.  
As pointed out in \citet{msc08}, there is a clear trend of spectral flattening/steepening towards lower/higher frequencies among this sample of local infrared-bright galaxies.    
In the bottom panel, only sources primarily powered by star formation (i.e., 6.2\,$\mu$m PAH EQW $\geq$ 0.27\,$\mu$m) are considered.  
Over plotted are expected spectral indices for star forming galaxies having a non-thermal spectral index of $\alpha \approx 0.83$ and 1.4\,GHz thermal fractions of $f_{\rm T}^{\rm 1.4GHz} \approx$5 (dashed line), 10 (dotted line), and 15\% (dot-dashed line) at 1.4\,GHz (see Figure \ref{fig:radspec}).  
%The dotted line indicates the spectral indices expected for a normal star-forming galaxy having a non-thermal spectral index of $\alpha \approx 0.83$ with a thermal fraction of $\approx$10\% at 1.4\,GHz (see Figure \ref{fig:radspec}).  
The solid line uses the expectation for the model with a 1.4\,GHz thermal fraction of 5\%, except sets the free-free optical depth to unity at 1\,GHz.  
This model seems to do a reasonable job fitting through the data points among the sample of star formation dominated sources.   
}
\label{fig:avgspx}
\end{figure}

\begin{figure}
\epsscale{1.1}
\plotone{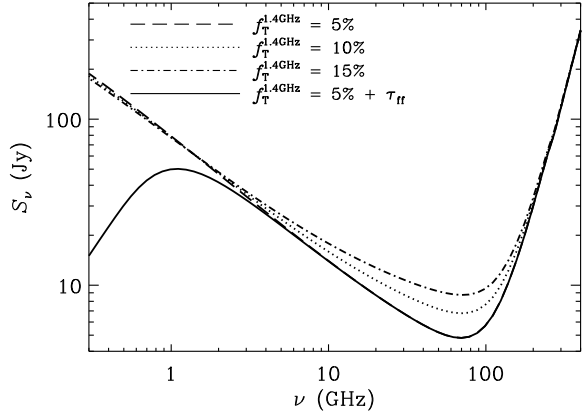}
\caption{Model radio-to-infrared galaxy spectra used for the comparison in Figure \ref{fig:avgspx}.  
The infrared ($8-1000\,\mu$m) flux is set to $F_{\rm IR} = 10^{-9}$\,W\,m$^{-2}$ and it is assumed that the source follows the far-infrared/radio correlation.  
The radio spectra are constructed following the prescription given in \citet{ejm09c}, resulting in a non-thermal spectral index of $\approx$0.83 with 1.4\,GHz thermal radio fractions of $f_{\rm T}^{\rm 1.4GHz} \approx$5 (dashed line), 10 (dotted line), and 15\% (dot-dashed line).  
The solid line is the same as the 5\% thermal fraction at 1.4\,GHz model, except that the free-free optical depth becomes unity at 1\,GHz.  
}
\label{fig:radspec}
\end{figure}

\section{Results}
%In a number of these infrared-bright starbursts, their high-frequency (i.e., $\ga 10\,$GHz) radio spectra are much steeper than expected for an increased thermal fraction, and in some cases, even show possible evidence for spectral steepening \citep{msc08, msc10, akl11b}.  
A number of the infrared-bright galaxies being investigated here are known to have high-frequency (i.e., $\ga 10\,$GHz) radio spectra that are much steeper than expected given that the thermal fraction should increase with frequency and work to flatten the spectra \citep[e.g.,][]{msc08, msc10, akl11b}. 
In the following section, correlations between various galaxy properties that may help to explain such steep, high-frequency radio spectral indices are investigated after removing potential AGN from the sample using the mid-infrared spectroscopic data.    
A number of physical considerations have already been put forth and discussed by \citet{msc08}, including those used to explain the high frequency turnover in the starburst galaxy NGC\,1569 by \citet{ute04}, and we refer the reader to these papers.  
To summarize, these physical processes include synchrotron aging, stochastic events such as radio hypernovae, rapid temporal variations in the star formation rate, and the escape of low energy CRs through convective transport.  
Each scenario has been found to be unsatisfactory for this sample of local starburst galaxies in a large part due to the extremely short \citep[i.e., $\sim10^{4}$\,yr;][]{jc91} radiative lifetimes estimated for CR electrons in such systems, which requires a near continuous injection of particles.  

%While these authors considered the case for which there may be a spectral break at $\sim$15\,GHz, strong evidence for such a turnover is still 
%Given that there is no strong evidence suggestion a high-frequency ``break" in the spectrum, we assume that each source has a single break at lower frequencies as the result of In the following analysis, we simply a

%\subsection{Identifying Potential AGN-dominated Sources}

\subsection{Radio Spectral Curvature}
%For this investigation, 
The radio spectra of each source are broken up into low-, mid-, and high-frequency bins, so the spectral curvature of each source can be crudely measured.   
The low frequency bin is defined as $\nu < 5\,$GHz, the mid-frequency bin spans $1\,{\rm GHz} < \nu < 10\,$GHz, and the high-frequency bin is defined as $\nu > 4$\,GHz.  
This allows the radio spectral index to be calculated using typically 3 or more data points in each frequency bin (see exact numbers in Table \ref{tbl-2}).  
The spectral index is estimated by an ordinary least squares fit over each frequency bin range, weighted by the photometric errors.  
Each spectral index is given in Table \ref{tbl-2} along with uncertainties from the fitting.  
Additionally given in Table \ref{tbl-2} is the average frequency, weighted by the signal-to-noise of each photometric data point, over which the radio spectral index was calculated.  

In the top panel of Figure \ref{fig:avgspx}, the median spectral indices of each bin are plotted as a function of the median frequency over which the spectral indices were calculated.  
The median frequencies of the low-, mid-, and high-frequency bins are $\approx$2, 4, and 12\,GHz, respectively.  
The horizontal error bar indicates the standard deviation in the average frequencies, while the vertical error bar is the error on the average spectral index.  
Average spectral indices are shown for the entire sample, as well as for stabursting and AGN-dominated systems, as defined by their 6.2\,$\mu$m PAH EQWs (see \S \ref{sec-data}).  

While a rather clear distinction between the average radio spectral indices measured between 1.49 and 8.44\,GHz (essentially the $\alpha_{mid}$ presented here) was pointed out by \citet{ejm13}, whereby AGN-dominated sources typically have a much flatter radio spectral index compared to starburst-dominated sources, the high-frequency radio spectral indices are statistically indistinguishable between both AGN- and starburst-dominated systems.  
Using the 6.2\,$\mu$m PAH EQW to split the sources up into starburst and AGN-dominated systems, the median high-frequency radio spectral index for starburst dominated systems is 0.78, with a standard deviation of 0.16.    
For the AGN-dominated systems,  the median spectral index is 0.71, albeit with a slightly larger scatter of 0.20.  
%Looking at $\alpha_{mid}$, starbursts and AGN have significantly different spectral indices, being 0.68 and 0.59 with standard deviations of 0.09 and 0.24, respectively.  
Thus, the steep spectral indices at $\sim 12\,$GHz do not seem to be the result of radio emission associated with an AGN.  
%Thus, based on these results, AGN-dominated system to have 

In the bottom panel of Figure \ref{fig:avgspx} the average spectral indices in each frequency bin are plotted for only those sources whose energetics are thought to be dominated by star formation based on having a 6.2\,$\mu$m PAH EQW $\geq 0.27\,\mu$m.  
The over plotted lines correspond to expectations based on 4 different radio spectra that are shown in Figure \ref{fig:radspec} to illustrate the change in radio spectral index as a function of frequency based on the increased thermal fraction towards higher frequencies for a normal star-forming galaxy.  
The radio-to-infrared models are based on the physical description given in \citet{ejm09c}, where the total infrared ($8-1000\,\mu$m) flux is set to $F_{\rm IR} = 10^{-9}$\,W\,m$^{-2}$ and the source is assumed to follow the far-infrared/radio correlation.  
The radio spectra are constructed following the prescription given in \citet{ejm09c}, resulting in a non-thermal spectral index of $\approx$0.83 with 1.4\,GHz thermal radio fractions of $f_{\rm T}^{\rm 1.4GHz} \approx 5$ (dashed line), 10 (dotted line), and 15\% (dot-dashed line).  
These 1.4\,GHz thermal fractions correspond to 30\,GHz thermal fractions of   $f_{\rm T}^{\rm 30GHz} \approx 33$, 51, and 63\%, respectively.  

As pointed out by \citet{msc08}, and also seen by \citet{akl11b} using new 33\,GHz VLA data, there is a clear trend of spectral flattening/steepening towards lower/higher frequencies among this sample of local, merger-driven starburst galaxies, which is clearly discrepant from the expectations of typical radio spectra for star-forming galaxies.    
However, by taking the model having a 1.4\,GHz thermal fraction of 5\%, modified by free-free absorption where the free-free optical depth is $\tau_{\rm ff} = (\nu/\nu_{\rm b})^{-2.1}$ and becomes unity at $\nu_{\rm b} = 1\,$GHz, the model largely fits the observed trend (solid line).    
%The dotted line illustrates the expected change in radio spectral index as a function of frequency based on the increased thermal fraction towards higher frequencies for a normal star-forming galaxy characterized by a non-thermal spectral index of $\alpha \approx 0.83$ and a thermal fraction of $\approx$10\% at 1.4\,GHz ($\approx$50\% at 30\,GHz).  

%\subsubsection{AGN-dominated systems}

\begin{figure}
\epsscale{1.1}
\plotone{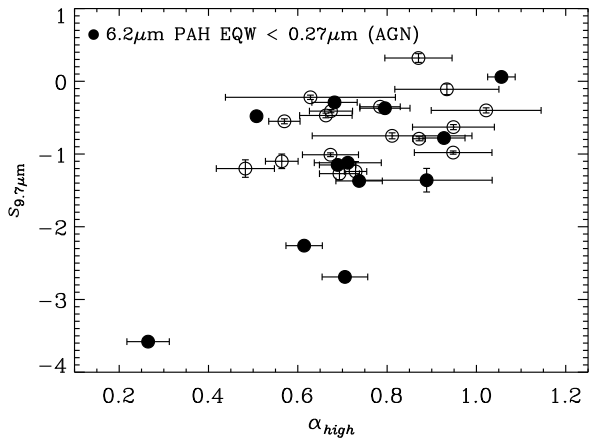}
\caption{The strength of the 9.7\,$\mu$m silicate feature plotted against the high-frequency radio spectral index (see Table \ref{tbl-2}). 
Sources categorized as being AGN dominated by their low 6.2\,$\mu$m PAH EQW are identified (filled circles).  
Unlike the radio spectral indices measured between 1.4 and 8.4\,GHz \citep{ejm13}, there does not appear to be any correlation between the silicate strength and the high-frequency radio spectral indices suggesting that the steep spectral indices at these frequencies may not be associated with the compactness of the starburst.  
}
\label{fig:spxhi-tau}
\end{figure}

\begin{figure}
\epsscale{1.1}
\plotone{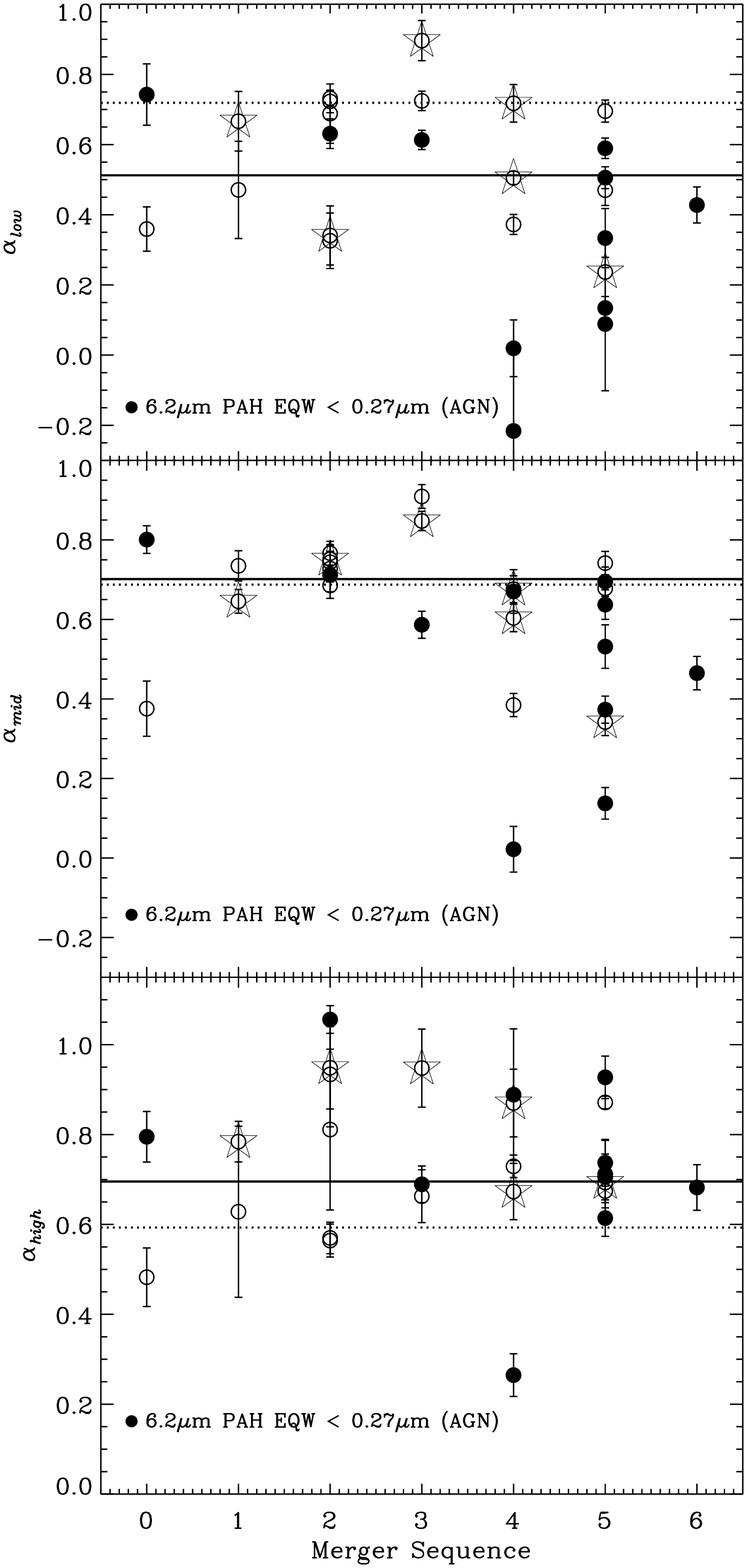}
\caption{The low-, mid-, and high-frequency radio spectral indices (see Table \ref{tbl-2}), plotted against merger stage (see \S\ref{sec-data}) in the top, middle, and bottom panels, respectively.  
The merger classification integers increase from isolated to post-merger systems.  
Sources categorized as being AGN dominated by their low 6.2\,$\mu$m PAH EQW are identified in each panel (filled circles). 
Those systems which are shown in Figure \ref{fig:maps} are marked with a star.  
The horizontal lines in each panel correspond to the expected spectral index at that frequency given the model radio spectra shown in Figure \ref{fig:radspec}.  
Only two models are shown; a normal star-forming galaxy having a 1.4\,GHz thermal fraction of 10\% (dotted line), and a star-forming galaxy with an original 1.4\,GHz thermal fraction of 5\% before being modified by free-free absorption in which the free-free optical depth becomes unity at 1\,GHz (solid line).  
Low- and mid-frequency radio spectral indices are found to be flat only for those sources classified as ongoing mergers with strongly disturbed morphologies and post-stage mergers.  
For the high-frequency radio spectral indices, the steepest radio spectral indices are found for systems classified as ongoing mergers with progenitors sharing a common envelope after removing AGN.  
%Both early- and post-stage mergers tend to have similar radio spectral indices, which are typical for normal isolated star-forming galaxies (i.e., $\alpha \approx 0.8$).  
}
\label{fig:MS-spx}
\end{figure}

\subsection{The High-Frequency Indices Compared to the Compactness of the Sources}
Recently, it has been shown that the flattening of the radio spectrum measured between 1.4 and 8.4\,GHz (i.e., essentially the spectral index in the mid-frequency bin presented here) among this sample of local infrared-bright starbursts increases with increasing 9.7\,$\mu$m silicate optical depth, 8.44GHz brightness temperature, and decreasing size of the radio source even after removing potential AGN \citep{ejm13}, supporting the idea that compact starbursts show spectral flattening as the result of increased free-free absorption \citep{jc91}.  
In Figure \ref{fig:spxhi-tau} the strength of the 9.7\,$\mu$m silicate feature is plotted against the high-frequency radio spectral index indicating the absence of any such trend.  
% that is observed when using the mid-frequency radio spectral index.  
Even after removing sources identified as harboring AGN due to their small 6.2\,$\mu$m PAH EQW, a trend is still not found.  
This indicates that the compactness of the starburst likely does not affect the steepness of the high-frequency radio spectrum.  

\subsection{The Radio Spectral Indices as a Function of Merger Stage}
As already shown by \citet{jc91}, there is a general trend among this sample of local starbursts in which more compact (i.e., higher surface brightness) sources tend to have flatter radio spectral indices, consistent with compact sources becoming optically thick at low (i.e., $\nu \la 5$\,GHz) radio frequencies.  
Another way one can illustrate this is by plotting the radio spectral index as a function of merger stage, which is done for the low-, mid-, and high-frequency radio spectral indices in Figure \ref{fig:MS-spx}.
The top, middle, and bottom panels plot radio spectral indices at frequencies of $\sim$2, 4, and 12\,GHz, respectively, against merger stage.  
The horizontal lines in each panel corresponds to the expected spectral index at that frequency given the model radio spectra shown in Figure \ref{fig:radspec}.  
Only two models are shown; a normal star-forming galaxy having a 1.4\,GHz thermal fraction of 10\%, and a star-forming galaxy with an original 1.4\,GHz thermal fraction of 5\% except that the free-free optical depth becomes unity at 1\,GHz.

In the top and middle panels, it seems that the flattest spectrum sources falling below what is expected for a normal star-forming galaxy, as indicated by the dotted line, are only found in ongoing and post-mergers with strongly disturbed morphologies.  % and post-stage mergers.  
The only outlier appears to be the non-merger IRAS\,F10173+0828, which has a flat spectrum at all frequency bins and hosts an OH megamaser \citep{ms87} along with a highly compact radio core \citep{lsl93}.  
This finding is robust even after removal of potential AGN as indicated by low 6.2\,$\mu$m PAH EQWs.  
However, this trend does not persist when the high-frequency spectral indices, measured at $\sim$12\,GHz, are plotted against merger sequence in the bottom panel.  
%(see bottom panel of Figure \ref{fig:MS-spx}).    
%Rather, it is found that, after removing potential AGN, the high-frequency spectral indices remain rather constant among pre- and post-stage mergers, while the steepest spectrum sources, falling well above what is expected for a normal star-forming galaxy as indicated by the dotted line, appear to be ongoing mergers in which galaxy nuclei are distinct, but share a common envelope and/or exhibit tidal tails as observed in their stellar light.  
%Rather, it is found that, after removing potential AGN, the high-frequency spectral indices appear rather constant across the full merger sequence.  
After removing potential AGN, the steepest spectrum sources, falling well above what is expected for a normal star-forming galaxy as indicated by the dotted line, are only found among those sources categorized as ongoing mergers in which galaxy nuclei are distinct, but share a common envelope and/or exhibit tidal tails as observed in their stellar light.  
In contrast, pre- and post-stage mergers do not seem to exhibit such steep spectral indices.  
Thus, the low-, mid-, and high-frequency spectral indices each appear to be sensitive to the exact stage of the merger.

\begin{figure}
\epsscale{1.1}
\plotone{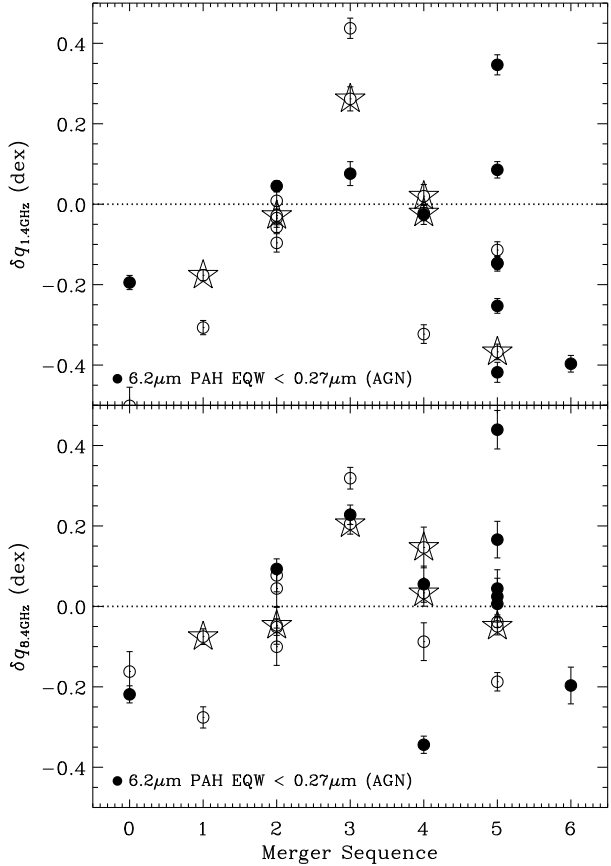}
\caption{{\it Top:} The difference between the observed and nominal logarithmic $F_{\rm FIR}$ to 1.4\,GHz flux density ratios (see \S\ref{sec:MS-q}), such that positive numbers indicate an excess of 1.4\,GHz radio continuum emission per unit star formation rate, assuming that the star formation rate is linearly proportional to the far-infrared emission, plotted against merger stage (see \S\ref{sec-data}; numbers increase from isolated to post-merger systems).  
{\it Bottom:} The same as the top panel, except using the difference between the observed and nominal $F_{\rm FIR}$ to 8.4\,GHz flux density ratios that, unlike the 1.4\,GHz data, are less affected by spectral flattening due to the increased free-free optical depth in the most compact starbursts.  
Sources categorized as being AGN dominated by their low 6.2\,$\mu$m PAH EQW are identified in each panel (filled circles).  
Those systems which are shown in Figure \ref{fig:maps} are marked with a star.  
In both panels, there seems to be a clear amount of excess radio emission relative to far-infrared emission among systems classified as ongoing mergers with progenitors sharing a common envelope.  
This distinction appears even more pronounced when potential AGN are removed.  
}
\label{fig:MS-q}
\end{figure}

\begin{figure*}
\epsscale{1.175}
\plotone{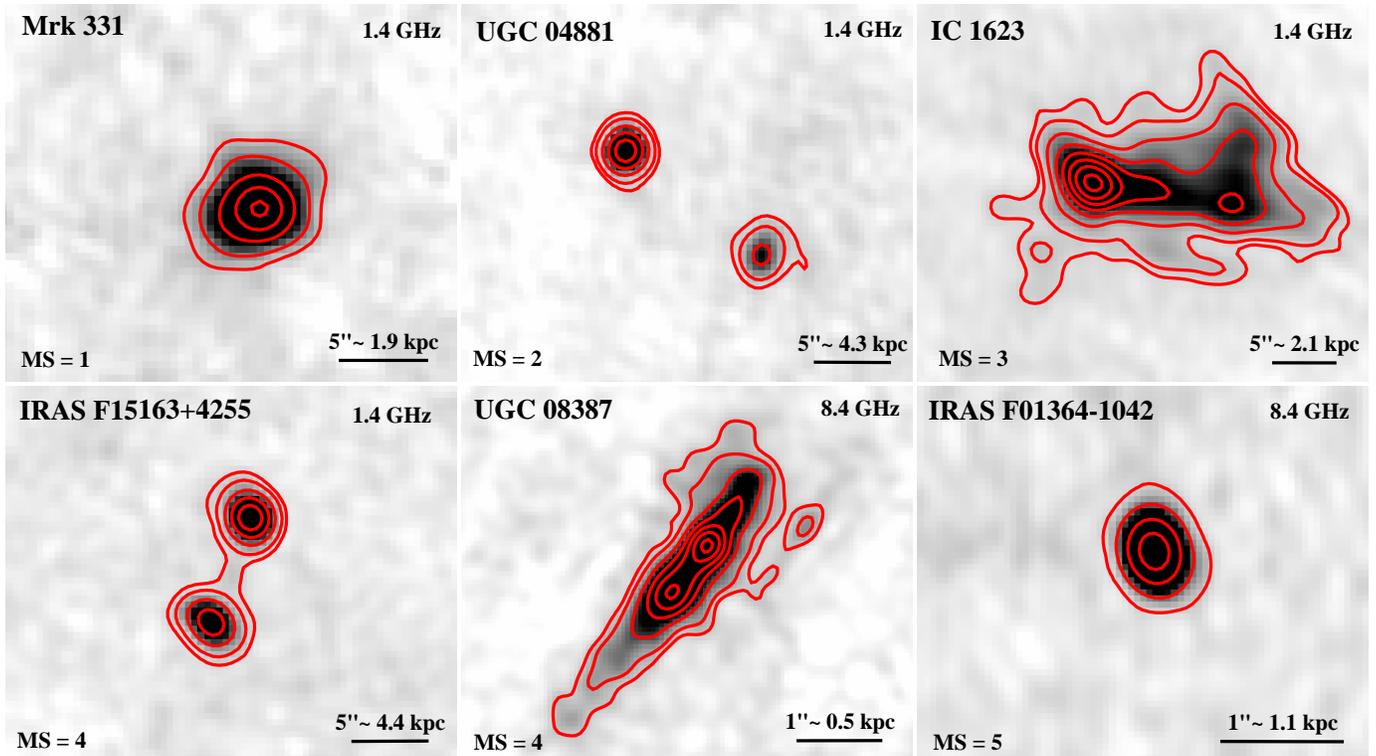}
\caption{Radio continuum maps and contours for 6 non-AGN systems, chosen to span the range of merger stages classified through {\it HST} imaging, are displayed (merger stage classifications are given in the lower left corner of each panel).  
For Mrk\,331, UGC\,04881, IC\,1623, and IRAS\,F15163+4255, 1.4\,GHz maps from \citet{chss90} are shown, all having resolutions of 1\farcs5 except for IC\,1623 (2\farcs1).     
However, for the more compact sources UGC\,08387 and IRAS\,F01364-1042, the higher resolution (0\farcs25) 8.4\,GHz maps from \citet{jc91} are shown.  
The contour levels start at 3 times the RMS noise of each map and are shown using a square-root scaling.    
For the pre-merger, Mrk\,331, its companion galaxy is not shown in the panel, as it is $\approx$2\arcmin away.   
}
\label{fig:maps}
\end{figure*}

\subsection{Far-Infrared-to-Radio Ratios as a Function of Merger Stage}
\label{sec:MS-q}
In Figure \ref{fig:MS-spx} it has been shown that sources having the steepest spectra appear to be associated with systems classified as ongoing mergers in which the progenitors are separable and share a common envelope or display strong tidal features.  
%While the index appears sensitive to the merger stage,  is interesting to see how the merger stage may affect the total radio continuum emission properties.  
To determine whether merger stage affects the total radio continuum emission, rather than just the spectral index, 
%In Figure \ref{fig:MS-q} 
the difference between the nominal and observed logarithmic $F_{\rm FIR}$/radio ratios at 1.4 and 8.4\,GHz are plotted in the top and bottom panels of Figure \ref{fig:MS-q}, respectively, against merger stage.  
The difference between the nominal and observed $F_{\rm FIR}$/radio ratios are defined as, 
\begin{equation}
\label{eq-deltaq}
\delta q_{\nu} = q_{0,\nu} - q_{\nu}, 
\end{equation}
such that a positive value indicates excess radio emission per unit far-infrared emission, and $q_{0,\nu}$ is the nominal $F_{\rm FIR}$/radio ratio for local star-forming systems.  
Under the assumption that the far-infrared emission is a good measure for the total star formation rate in each system, a large value of $\delta q_{\nu}$ indicates excess radio emission per unit star formation rate.  
At 1.4\,GHz, $q_{\rm 0,1.4GHz} \approx 2.34$\,dex with a scatter of 0.26\,dex among $\sim$1800 galaxies spanning nearly 5 orders of magnitude in luminosity \citep[e.g.][]{yrc01}, among other properties (e.g. Hubble type, far-infrared color, and $F_{\rm FIR}$/optical ratio).  
Assuming a typical radio spectrum, with a spectral index of $\alpha \approx 0.8$ \citep{jc92}, this translates into a nominal $F_{\rm FIR}$ to 8.4\,GHz flux density ratio of $q_{\rm 0,8.4GHz} \approx 2.94$\,dex.  

In the top panel of Figure \ref{fig:MS-q}, it can be shown that galaxies having the largest 1.4\,GHz excesses appear to be ongoing mergers that share a common envelope before and after removal of potential AGN.  
However, there is a good deal of scatter in this figure, and there are a number of sources for which significant excesses of far-infrared emission are observed relative to the measured 1.4\,GHz flux densities.  
Since the $q_{\rm 1.4GHz}$ ratios are clearly affected by free-free absorption at low frequencies in the most compact starbursts \citep{jc91}, it may be better to see if such trend persists using emission at a higher frequency, where free-free absorption is considered to be largely negligible, such as 8.4\,GHz.  
This is shown in the bottom panel of Figure \ref{fig:MS-q}, where $\delta q_{\rm 8.4GHz}$ is plotted against merger stage and the exact same behavior is found, albeit with a much smaller dispersion per merger sequence bin.  
Excluding potential AGN, there appears to be a trend in which sources classified as ongoing mergers that share a common envelope have excess radio emission per unit star formation rate relative to both early and post-stage mergers, which exhibit nominal $F_{\rm FIR}$/radio ratios.  
Thus, similar to radio spectral indices, the far-infrared-to-radio ratios appear to be sensitive to merger stage.

\section{Discussion}
As previously stated, a number of physical scenarios to explain the steep radio spectral indices among this sample of local starburst galaxies have been proposed in the literature, particularly by the original study of \citet{msc08}.  
While simple arguments to produce steep spectra via increased synchrotron and inverse Compton losses appear to fall short, given that the most compact starbursts are not the sources that exhibit the steepest spectra, an alternative explanation is proposed.  
It has been know for some time that, when caught in the act, merging systems can create bridges of synchrotron emission that stretch between the progenitor galaxies, whose brightness contours resemble stretching strands of ``taffy" \citep{jc93}.   
The two most well-known cases of ``taffy" galaxies are the face-on colliding systems UGC\,12914/5 \citep{jc93} and UGC\,813/6 \citep{jc02}.  

%The radio continuum bridge connecting the ÔÔ taffy ÕÕ gal- axies UGC 12914 and UGC 12915 (Condon et al. 1993, hereafter Paper I) provides an independent observational diagnostic for galaxy collisions. 
The space between the spiral galaxy pairs  UGC\,12915/5 and UGC\,813/6 are filled with synchrotron emission, implying that they must contain both relativistic electrons and magnetic fields.  
The magnetic field of the bridge is presumably stripped from the merging spiral galaxy disks as they interpenetrated during a recent collision, however there is some debate as to whether the relativistic electrons in the bridge have escaped the merging spirals, as originally suggested by \citet{jc93}, or are rather a new population of relativistic electrons that have been diffusively accelerated in a shock associated with a gas dynamical interaction during the merger \citep{ute10}.  
As discussed below, if such a scenario is also responsible for the properties of the starbursting mergers investigated here, the latter explanation appears more appealing.  
%{\bf RWD and state what we prefer here... A preference for which scenario best fits these starbursts is discussed later. }

Other prominent features of the ``taffy "systems include: 
(1) An exceptionally steep spectrum located at the center of the radio bridges (i.e., $\alpha \approx 1.3$, which is  roughly $\Delta\alpha= 0.5$ steeper than the spectral indices of the  galaxy disks). 
(2) A ratio of $F_{\rm FIR}$/radio emission for the entire system that is significantly lower than that of normal star-forming galaxies, but typical for the individual spiral galaxy disks.  
(3) A significant fraction of H{\sc i} stripped by the collisions that resides between the spiral disks. 
(4) A significant amount of molecular gas in the bridge \citep{jb03} whose physical conditions are comparable to those in the diffuse clouds of the Galaxy \citep{zhu07}.   
(5) A small amount of warm \citep[i.e., $5-17\,\mu$m;][]{tj99} and cold \citep[i.e., 450 and 850\,$\mu$m;][]{zhu07} dust located in the bridge, indicating a low amount on ongoing star formation.  
(6) Rotational lines of H$_{2}$ emission that dominate the mid-infrared spectrum and appear strongest near the center of the bridge \citep{bwp12}.  
%These unusual features were interpreted (Paper I, Paper II) as the expected outcome of a face-on collision between two (formerly) normal spiral galaxies.

\subsection{``Taffy"-Like Starbursts}
Among the merging starbursts investigated here, those having the steepest high-frequency ($\nu \sim 12$\,GHz) radio spectra, and significant excesses of radio emission relative to their observed far-infrared emission, appear to be galaxies in an ongoing merger and either share a common stellar envelope or have significant tails.  
Both of these radio characteristics are also found in ``taffy" systems, whose steep spectra and excess radio emission per unit star formation rate arises from the radio continuum emission in the bridge connecting the merging galaxies.  
Given that simple physical arguments based on the escape and radiative cooling times of CR electrons to explain these properties among local compact starburst fail, a taffy-like merger scenario appears to be an appealing way to explain the low $F_{\rm FIR}$/radio ratios and steep high-frequency spectra.  

In Figure \ref{fig:maps}, radio continuum maps are shown for 6 galaxies that largely span the full merging sequence and do not appear to harbor buried AGN based on their measured 6.2\,$\mu$m PAH EQWs.  
Each system has been highlighted as a star in Figures \ref{fig:MS-spx} and \ref{fig:MS-q}.  
Mrk\,331 is a pre-merger, exhibiting a symmetric 1.4\,GHz radio continuum morphology, as well as a typical radio spectrum and $F_{\rm FIR}$/radio ratio.  
UGC\,04881 is characterized as an ongoing merger with separable progenitor galaxies.  
IC\,1623 is optically classified as an ongoing merger whose progenitors share a common envelope, which is consistent with its 1.4 and 8.4\,GHz radio morphologies.  
It has a high-frequency radio spectral index of $\approx$0.95, and a factor of 1.6 excess radio emission relative to what is expected given the far-infrared/radio correlation.  
The excess radio emission appears easily explained by summing the diffuse synchrotron emission that is not associated with the colliding star-forming galaxy disks.      
IRAS\,F15163+4255 is classified (by the {\it HST} data) as an ongoing merger with two nuclei and a prominent tidal tail.  
This source has very weak 1.4 and 8.4\,GHz radio continuum emission that appears to form a bridge between the two progenitor galaxies, keeping it right on the far-infrared/radio correlation.  
This source also exhibits a somewhat steep high frequency radio spectral index of $\approx$0.84.  
Like IRAS\,F15163+4255, UGC\,08387 is also classified as an ongoing merger harboring two nuclei and a prominent tidal tail.    
Given that the progenitor galaxies are not separated by a large angular extent, the high (0\farcs25) 8.4\,GHz radio continuum map of \citet{jc91} is necessary to resolve both components.  
Similar to IC\,1623, the 8.4\,GHz morphology is consistent with its optical merger classification, as there are tidal tails of synchrotron emission.  
UGC\,08387 has a high-frequency radio spectral index of $\approx0.70$ and 40\% excess radio emission at 8.4\,GHz than expected giving its far-infrared flux.  
Finally, the post-stage, ultra-compact, merger IRAS\,F01364+1042 is shown, also requiring the use of the high-resolution 8.4\,GHz radio map in which their appears to be a single source having seemingly undisturbed radio continuum contours.  
This system has an extremely flat low- and mid-frequency radio spectral index, presumably due to free-free absorption occurring in the compact starbursts, but a rather normal high-frequency spectral index.  
Consequently, its $q_{\rm 1.4GHz}$ value is significantly larger than average, unlike its rather typical $q_{\rm 8.4GHz}$ value.

\begin{figure}
\epsscale{1.2}
\plotone{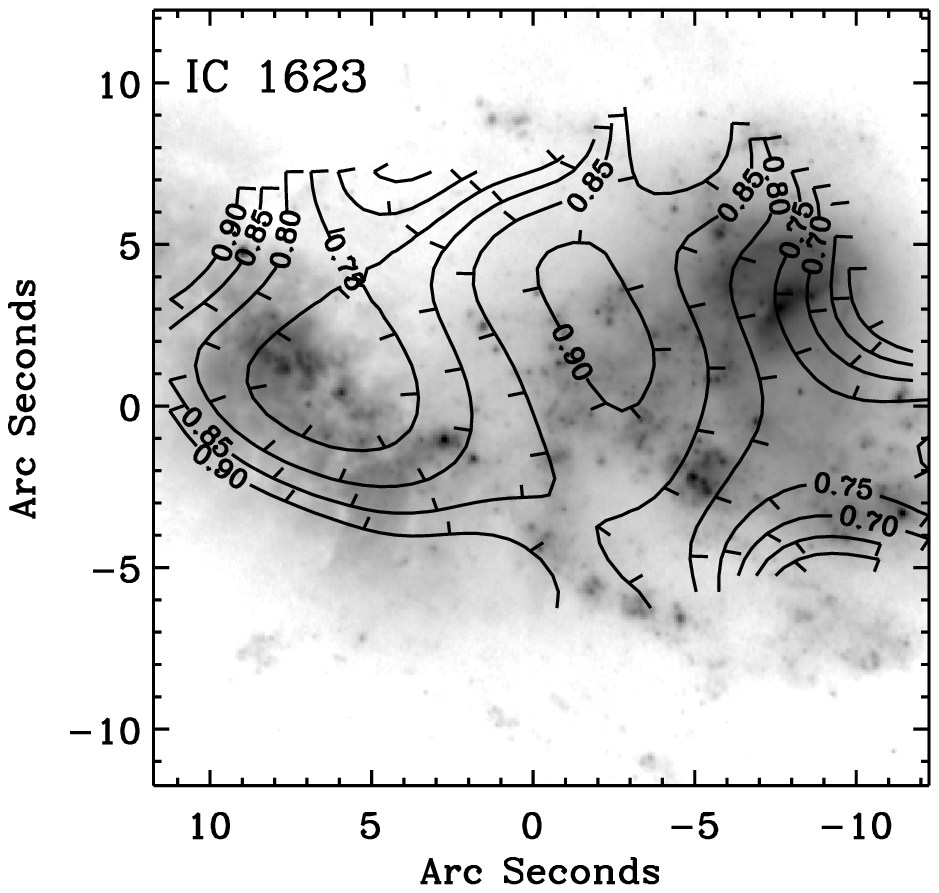}
\caption{
Spectral index (1.4 to 8.4\,GHz) contours overlaid on an {\it HST}/ACS image of IC\,1623 taken with the F184W ($I$-band) filter \citep{dck13}.  
The resolution of the spectral index contours is 4\farcs7, set by the natural-weighted 8.4\,GHz image from \citet{msc08}.  
The radio spectral index is flattest on the infrared-bright starburst core to the east ($\alpha^{\rm 8.4GHz}_{\rm 1.4GHz} \approx 0.7$), which is still typical for normal star-forming galaxies.  
The spectral index along the much less dust-obscured western galaxy disk is also normal, being $\alpha^{\rm 8.4GHz}_{\rm 1.4GHz} \approx 0.75$.  
However, similar to the ``taffy" galaxies, the radio spectrum steepens significantly along the radio continuum bridge connecting the galaxy pair, peaking at $\alpha^{\rm 8.4GHz}_{\rm 1.4GHz} \approx 0.92$ at the midpoint between the interacting disks.   
}  
\label{fig:spxmap}
\end{figure}

\begin{figure}
\epsscale{1.1}
\plotone{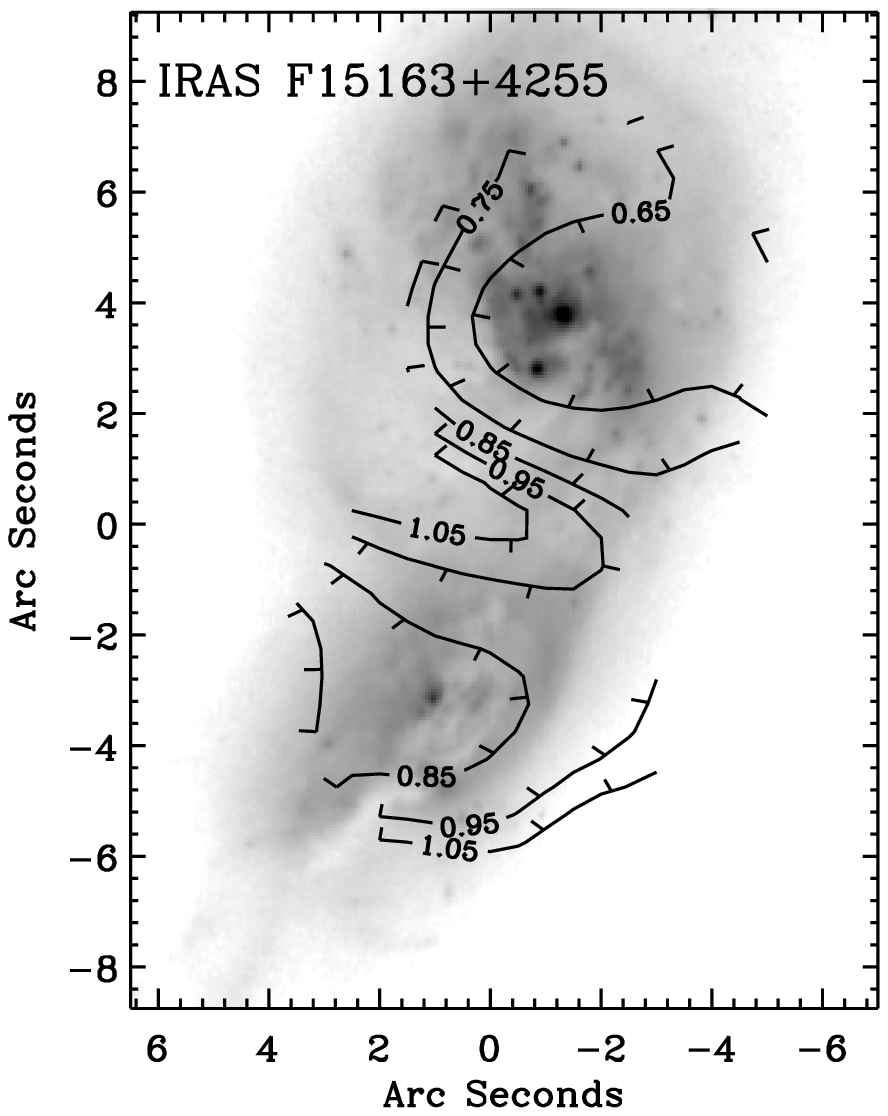}
\caption{
Spectral index (1.4 to 8.4\,GHz) contours overlaid on an {\it HST}/ACS image of IRAS\,F15163+4255 taken with the F184W ($I$-band) filter \citep{dck13}.  
The resolution of the spectral index contours is 3\farcs1, set by the natural-weighted 8.4\,GHz image from \citet{msc08}.  
The radio spectral index of the southern galaxy is $\alpha^{\rm 8.4GHz}_{\rm 1.4GHz} \approx 0.8$, which is still typical for normal star-forming galaxies.  
The spectral index of the brighter northern galaxy is significantly flatter, being $\alpha^{\rm 8.4GHz}_{\rm 1.4GHz} \approx 0.6$.  
Similar to the ``taffy" galaxies, the radio spectrum steepens significantly along the, in this case very weak, radio continuum bridge connecting the galaxy pair, peaking at $\alpha^{\rm 8.4GHz}_{\rm 1.4GHz} \approx 1.1$ at the midpoint between the interacting disks.   
}  
\label{fig:spxmap2}
\end{figure}

Focusing in on IC\,1623, 1.4 to 8.4\,GHz (4\farcs7 resolution) spectral index contours are overlaid on an $I$-band (F814W) {\it HST}/ACS image in Figure \ref{fig:spxmap}.  
%In Figure \ref{fig:spxmap}  
The radio spectral indices are found to be rather typical relative to normal star-forming galaxies for both merging galaxy disks (i.e., $\alpha^{\rm 8.4GHz}_{\rm 1.4GHz} \sim 0.7$).    
The eastern galaxy disk hosts the most deeply embedded star formation, and the radio spectral index clearly flattens right on the peak of the starburst.  
Similar to the ``taffy" systems, the radio spectral index is found to steepen significantly along the radio continuum bridge connecting the galaxy pair, peaking roughly at the midpoint between the two galaxy disks with $\alpha^{\rm 8.4GHz}_{\rm 1.4GHz} \approx 0.92$.  
Furthermore, like the ``taffy" systems, high-resolution $^{12}$CO observations show a significant amount of molecular gas located in the overlap regions connecting the two galaxy nuclei \citep{my94,di04}, thus providing material for the galactic magnetic fields to be anchored as they stretch during the merger process. 
A similar situation is observed by plotting the 1.4 to 8.4\,GHz (3\farcs1 resolution) spectral index contours on the $I$-band (F814W) {\it HST}/ACS image of IRAS\,F15163+4255 in Figure \ref{fig:spxmap2}.  
While the northern and southern galaxy nuclei appear to have significantly different spectral indices, peaking around  $\alpha^{\rm 8.4GHz}_{\rm 1.4GHz} \approx 0.6$  and $\alpha^{\rm 8.4GHz}_{\rm 1.4GHz} \approx 0.8$ , respectively, there is a strong steepening of the indices towards the emission connecting the two galaxy nuclei peaking at $\alpha^{\rm 8.4GHz}_{\rm 1.4GHz} \approx 1.1$ at the midpoint between the interacting disks. 
%In Figure \ref{fig:spxmap}  

Explaining the steep radio spectral indices and lower than average $F_{\rm FIR}$/radio ratios as the result of ongoing mergers seems to be fairly well supported by the data. 
%, and in fact a very 
To date, it has been hard to explain the steep high-frequency spectra in this sample of starbursts given that the rapid ($\sim10^{4}$\,yr) synchrotron and inverse Compton cooling times render diffusion and escape losses to be negligible, and thus cannot work to steepen the spectra.    
In fact, escape as an explanation is in disagreement with the data, given that the loss of CR electrons would effectively lower the total synchrotron power, whereas galaxies with steep high-frequency spectral indices can have low $F_{\rm FIR}$/radio ratios.  
By creating a magnetized medium through the dynamical interaction of the merging galaxy pairs, similar to that of the taffy systems, provides a location where the CR electrons can radiatively cool without the continuous injection and acceleration of new particles.  

Given the rapid cooling times due to radiative losses in the compact starbursts embedded in the merging disks, it seems unlikely that the relativistic electrons associated with synchrotron bridges and/or tidal tails were injected and accelerated in the starburst itself, and have propagated to such distances.  
As an example, the distance to the midpoint between the two 1.4\,GHz ``hot spots" in the merging disks of IC\,1623 is $\approx$5\arcsec, which projects to a linear distance of $\approx$2\,kpc at a distance of 85.5\,kpc.  
This distance is actually smaller than the $\approx$15\arcsec (6\,kpc) separation between the two galaxy nuclei.  
Following the description for propagation and physical processes responsible for CR electron energy losses given in \citet{ejm09c}, it can be shown that the time for 1.4\,GHz emitting electrons to reach this midpoint is significantly longer than their radiative lifetime.  
Assuming random walk diffusion, characterized by an energy-dependent diffusion coefficient, the propagation time is $\sim 2.6\times 10^{7}$\,yr.  
The radiative lifetime for CR electrons emitted within a 2\farcs5 radius of the western galaxy, which likely includes the bulk 1.4\,GHz emission associated with star formation in that disk, is $\sim2.9\times10^{5}$\,yr, nearly two orders of magnitude shorter.  

Additionally, the time that the merger is expected to spend in either of these classifications is no more than a few times $\sim10^{7}$\,yr \citep{haan11}, which is roughly an order of magnitude longer than the synchrotron cooling times of CR electrons that may populate a bridge, assuming the magnetic field stays in equipartition with the disk it is being pulled out of, and that the radiation field and ISM density leads to negligible inverse Compton and bremsstrahlung cooling, respectively.  
Thus, it appears more likely that charged particles in such regions have been diffusively accelerated via shocks associated with the available mechanical energy of the merger \citep[e.g.,][]{ute10}.  
The speed of the galaxy collisions are a few $\sim$100\,km\,s$^{-1}$ (e.g., G. Privon et al. 2013, in preparation), and therefore super-Alfv\'{e}nic for typical ISM Alfv\'{e}n speeds a few $\sim$10\,km\,s$^{-1}$.  
The Alfv\'{e}nic Mach number is much larger than unity,  at the order $M_{\rm A}^{2} \sim \mathcal{O}(100) \gg 1$, allowing for shocks that can diffusively accelerate CR particles.  
This is similar to the Alfv\'{e}nic Mach numbers within the Sedov-Taylor expansion phase of a SNRs.  

This explanation for diffusive acceleration via shocks associated with the merger is additionally supported by integral field spectroscopic observations of infrared-bright mergers.  
For the case of IC\,1623, \citet{jrich11} have shown that optical emission line diagnostic ratios indicate the presence of widespread shock excitation induced by ongoing merger activity.  
The energy associated with the shocks in the interacting regions between the two galaxy disks is estimated to be $\sim 4\times10^{42}$\,erg\,s$^{-1}$ based on the amount of H$\alpha$ line emission having widths $\gtrsim$100\,km\,s$^{-1}$ in the interacting region \citep[i.e., $80\times L_{\rm H\alpha}$ where $L_{\rm H\alpha} \sim 5 \times 10^{40}$\,erg\,s$^{-1}$;][]{jrich11,jrich10}.  
This value is nearly a factor of $\sim1.5\times10^3$ times larger than the total 8.4\,GHz luminosity of the source ($\nu L_{\rm 8.4GHz} \sim 3\times10^{39}$\,erg\,s$^{-1}$), suggesting that, for a proton-to-electron ratio of $\sim$100 in this GeV energy range, $\sim$4\% of the total mechanical luminosity from the shock needs to go into accelerating CRs to explain a factor of $\sim$2 extra radio emission.  
This is much less than the typical $10-30$\% efficiency of particle acceleration in SNRs \citep[e.g.,][]{berz97,kl05,dcap10}.

\subsection{The Case for a Steeper Injection Spectrum in Dense Starbursts}
While it is argued here that the most likely explanation for both the steep high-frequency radio spectra and ``excess" radio emission in this sample of local starbursts arises from synchrotron bridges and tails associated with the stage of the merger, another explanation that has not been currently explored is a systematic change (steepening) in the injection spectrum in this sample of starbursts.  
Naively, such an scenario does not seem that implausible given the ISM conditions in dense starbursts.  

The efficiency in which CRs are accelerated in SNRs plays a significant role in determining the synchrotron emissivity from galaxies.  
One could imagine a scenario in which the adiabatic phase of supernovae is halted as it expands into the ambient medium, thus reducing the amount of energy lost to adiabatic expansion leaving ``extra" energy that could be used in the acceleration of CRs, thereby increasing the total synchrotron emission per unit star formation rate relative to normal galaxies.  
For example, the modeling of \citet{ed91,ed00} shows that the total acceleration efficiency for CRs increases from $\sim$0.15$E_{\rm SN}$ to $\sim$0.25$E_{\rm SN}$, where $E_{\rm SN}=10^{51}$~erg is the total explosion energy of the SNe, when the external ISM density is increased from $\sim$1~cm$^{-3}$ to $\sim$10~cm$^{-3}$.  
Additionally, an increase in the injection efficiency can work to steepen the CR injection spectrum \citep{dcap12}.  
The ISM densities of starbursts are typically much larger (e.g., $n_{\rm ISM} \sim 10^{4}$~cm$^{-3}$), thus the evolution of SNRs will likely be different and may work to increase the synchrotron emissivity in such galaxies through an increased efficiency in particle acceleration, as well as result in steeper radio spectra due to a steeper initial injection spectra.

However, given that the densest starbursts, which exhibit the flattest low- and mid-frequency spectral indices are associated with late/post-stage mergers, and also have typical high-frequency radio spectral indices as well as normal $F_{\rm FIR}$/radio flux density ratios, this explanation seems less likely.  
This additionally suggests that the excess radio emission in starbursts as a result of increased secondary electrons may not be a likely explanation \citep[e.g.,][]{ejm09c,ltq10}, since the densest starbursts (i.e., post mergers) provide the environment for which secondary production should be the most efficient.  
For example, in a dense starburst,  having a much larger ISM density, the cross-section for collisions between CR nuclei and the interstellar gas is increased, thus increasing the number of $e^{\pm}$'s for a fixed primary nuclei/electron ratio, which may actually dominate the diffuse synchrotron emission.    
Yet, the densest, post-merger starbursts do not show evidence for excess radio emission per unit star formation rate like those systems in which the progenitors are still clearly separated and exhibit a non-negligible amount of diffuse radio emission in features such bridges and tails.

\subsection{An Explanation for Low FIR/Radio Ratios in High-z SMGs?}
A significant fraction of submillimeter galaxies (SMGs) detected at redshifts between $2 \la z \la 4$ similarly show excess (i.e., a factor of $\ga$3) radio emission relative to their total far-infrared emission and a nominal $F_{\rm FIR}$/radio ratio \citep[e.g.,][]{ak06, ev07, pc08,ejm09b, ed09,ed09b,kekc09, kk10,vsmol11}.  
Although, it is worth pointing out that this is not true for all samples of SMGs at these redshifts  \citep[e.g.,][]{sc10}.  %exhibit discrepant $F_{\rm FIR}$/radio ratios
AGN provide a likely explanation for the excess radio emission in such sources given that a number are detected in hard ($2.0-8.0$\,keV) X-rays \citep{da05}, 
%Analysis of the X-ray data on SMGs also indicates the presence of an AGN in most SMGs (Alexander et al. 2005). Thus, it is very likely that some of the radio emission arises from the nuclear black hole as can be seen by the fact that a number of the most deviant qIR values are in fact detected in the hard band (2.0Ð8.0 keV) X-rays.
however, it is possible that those sources without evidence for AGN may exhibit excess radio emission associated with being involved in an ongoing merger.  
% to being involved in an ongoing merger.  
%it may be that a number of such sources exhibit excess radio emission due to being in ongoing mergers.  

At such cosmological distances, it is currently unclear what fraction of SMGs are major mergers rather than isolated disks, but the morphologies for a number of resolved sources seem to suggest major mergers that are driving intense bursts of star formation \citep[e.g.,][]{sc03a,jah13}.
With the Atacama Large Millimeter/submillimeter Array (ALMA)  now online, resolving these dusty starbursts into individual components is becoming easier \citep[e.g.,][]{ydh13, jah13}.  
For instance, the lensed star-forming galaxy SPT-S\,053816-5030.8, at a redshift of  $z = 2.783$, has a radio spectrum that appears to flatten ($\alpha \approx 0.18$) towards low frequencies, while having a rather steep ($\alpha \approx 0.76$) spectrum at high frequencies \citep{ma13}, similar to what is seen among the local compact starbursts investigated here.  
%Correcting for the spectral flattening, assuming it is caused by free-free absorption in the dense starburst, the total 1.4\,GHz radio flux density of this system appears to be a factor of $\sim$2 larger than what is expected for its measured far-infrared flux and an average $F_{\rm FIR}$/radio ratios.  
While the estimated (rest-frame) $q_{\rm 1.4GHz}$ value for this source is slightly larger than the local average value, consistent with expectations given the spectral flattening at lower frequencies assuming optically-thick free-free emission, the $q_{\rm 8.3GHz}$ value is  $\approx$0.36\,dex smaller than the local average value, implying excess radio emissions per unit star formation rate.  

New imaging at 350\,GHz using ALMA along with lens modeling \citep{ydh13} suggests that this SMG is in fact composed of two galaxies, one of which is a compact source that dominates the far-infrared emission.  
Both the radio continuum properties and 350\,GHz morphology suggests that the source is consistent with being powered by merger-driven star formation as observed in local (U)LIRGs.  
Thus, at least for this SMG, excess non-thermal radio emission associated with the merger may provide a natural explanation for its steep radio spectral index at high frequencies and presumably excess radio emission per unit star formation rate relative to normal star-forming galaxies rather than requiring the need to invoke cosmic conspiracies \citep[e.g.,][]{lt10}.  
Having better multifrequency radio data for a large number of these high-redshift SMGs, to see if their spectra is also steep at high frequencies, may help to explain exactly why this population of starbursting galaxies appear to have significantly more radio emission per unit star formation rate than expected.  

\section{Conclusions}
An examination of the radio continuum properties for a sample of local infrared-bright starbursts has been investigated against their merger classification.  
This was done to shed light on the curious nature of the radio spectra among such sources, specifically those having steeper than expected radio spectra observed at frequencies $\sim 12$\,GHz as pointed out by \citet{msc08}.  
The main conclusions from this investigation can be summarized as follows:

\begin{enumerate}
\item{Sources categorized as starbursts and AGN via their 6.2\,$\mu$m PAH EQWs have similar high-frequency ($\nu \sim 12$\,GHz) radio spectral indices, suggesting that the steep spectra among some sources at these frequencies are not the result of radio emission associated with an AGN. 
}

\item{
Sources having the steepest radio spectral indices at $\sim$12\,GHz, while also appearing to be powered primarily by star formation as indicated by their 6.2\,$\mu$m PAH EQWs, appear to be classified as ongoing mergers in which the progenitors are still separated and either share a common envelope or show significant tidal tails in their stellar light.  
Similarly, these same galaxies also exhibit excess radio emission relative to what is expected given their observed far-infrared emission and the tight far-infrared/radio correlation, suggesting that there is excess radio emission not associated with ongoing star formation activity.  
The combination of these observations leads to a picture in which the steep high-frequency radio spectral indices and excess radio emission arises from radio continuum bridges and tidal tails in which a new population of relativistic electrons have been accelerated and can radiatively cool producing a steep spectrum.  
Such a scenario is consistent with high-resolution radio morphologies of the sources as a function of merger stage, as well as the radio spectral index map for the merging galaxy pairs IC\,1623 and IRAS\,F15163+4255.  
}

\item{
Among all sources whose energetics are thought to be dominated by star formation given their 6.2\,$\mu$m PAH EQWs, their average radio spectral indices at $\sim$2, 4, and 12\,GHz appear to be fairly well fit by a model radio spectrum having a non-thermal radio spectral index of $\approx$0.83, a 1.4\,GHz thermal fraction of $\approx$5\% (corresponding to $\approx$33\% at 30\,GHz) modified by free-free absorption where the free-free optical depth becomes unity at $\approx$1\,GHz.  
}

\end{enumerate}

%With the Karl G. Jansky VLA now online, we have reached a point where it is feasible to significantly improve upon the type of analysis presented in this paper by conducting higher frequency radio continuum surveys (e.g., $\ga$10\,GHz) of local starbursts, probing $\la$100\,pc scales \citep[e.g.,][]{akl11b}.  
%For instance, by having resolved sizes along with better spectral coverage, to identify the cutoff frequency at which the free-free optical depth equals unity, one can begin to characterize additional physical characteristics of the starbursts such as their emission measures and electron densities.  
%Additionally, we are likely to see a new generation of high-frequency, deep-field radio surveys that compliment existing 1.4\,GHz data, while providing sub-arcsecond resolution of distant dusty starbursts.  
%Such surveys will need to rely on investigations such as these to interpret their observations.  

\acknowledgements
We thank the anonymous referee for useful comments that helped to significantly improve the content and presentation of this paper.
%We thank J.H.~Howell, T.~D\'{i}az-Santos, and V. Charmandaris for useful discussions.  
E.J.M. thanks S. Stierwalt, S. Haan, J. A. Rich, G. C. Privon, L. Armus, L. Barcos, A.K. Leroy, and P.A. Appleton, for useful discussions, M. Clemens for providing reduced X-band maps, and D.-C. Kim and A.S. Evans for providing their reduced {\it HST} images. 
E.J.M. is also grateful to J.J. Condon for giving the paper a careful reading and providing useful comments.  %, and thanks M. Clemens, for providing his reduced X-band maps.   
E.J.M. acknowledges the hospitality of the Aspen Center for Physics, which is supported by the National Science Foundation Grant No. PHY-1066293.  
The National Radio Astronomy Observatory is a facility of the National Science Foundation operated under cooperative agreement by Associated Universities, Inc.
This work is based in part on observations made with the {\it Spitzer} Space Telescope, which is operated by the Jet Propulsion Laboratory, California Institute of Technology under a contract with NASA.  
This research has made use of the NASA/IPAC Extragalactic Database (NED) which is operated by the Jet Propulsion Laboratory, California Institute of Technology, under contract with the National Aeronautics and Space Administration.

%\bibliography{/Users/emurphy/libs/bibtexref/master_ref}
\bibliography{aph.bbl}

\end{document}

%% file: tbl-1.tex
\begin{deluxetable*}{lcccccccc}
\tablecaption{Radio and Infrared Properties of the Sample Galaxies \label{tbl-1}}
\tabletypesize{\scriptsize}
\tablewidth{0pt}
\tablehead{
\colhead{Galaxy}  & \colhead{Dist.\tablenotemark{a}} & \colhead{$L_{\rm IR}/10^{11}$\tablenotemark{a}} & \colhead{$F_{\rm FIR}/10^{-13}$\tablenotemark{b}} & \colhead{$q_{\rm 1.4GHz}$\tablenotemark{c}} &  \colhead{$q_{\rm 8.4GHz}$\tablenotemark{c}} &\colhead{$6.2\mu$m EQW\tablenotemark{d}} & \colhead{$s_{9.7\micron}$\tablenotemark{d}} & \colhead{Merger Stage\tablenotemark{e}}\\
\colhead{} & \colhead{(Mpc)} & \colhead{($L_{\sun}$)} & \colhead{(${\rm W\,m^{-2}}$)} & \colhead{(dex)} & \colhead{(dex)} & \colhead{($\mu$m)} & \colhead{} &  \colhead{}
}
\startdata
             NGC\,34&   84.1&   3.09&   7.69&   2.49&   3.13&     0.45/SF                 &       -0.79                 &           5\\
            IC\,1623&   85.5&   5.13&  11.18&   2.08&   2.74&     0.30/SF\tablenotemark{f}&       -0.98\tablenotemark{f}&           3\\
       CGCG\,436-030&  134.0&   4.90&   4.70&   2.40&   2.99&     0.35/SF                 &       -1.10                 &           2\\
   IRAS\,F01364-1042&  210.0&   7.08&   3.02&   2.71&   2.99&     0.39/SF                 &       -1.27                 &           5\\
   IRAS\,F01417+1651&  119.0&   4.37&   5.74&   2.58&   2.89&     \nodata                 &     \nodata                 &           3\\
           UGC\,2369&  136.0&   4.68&   3.94&   2.33&   2.90&     0.57/SB                 &       -0.11                 &           2\\
   IRAS\,F03359+1523&  152.0&   3.55&   2.88&   2.60&   2.84&     \nodata                 &     \nodata                 &           3\\
           NGC\,1614&   67.8&   4.47&  14.62&   2.45&   2.98&     0.61/SB                 &       -0.41                 &           5\\
   IRAS\,F05189-2524&  187.0&  14.45&   5.89&   2.74&   3.14&    0.03/AGN                 &       -0.29                 &           6\\
           NGC\,2623&   84.1&   4.37&  11.02&   2.49&   2.92&    0.27/AGN                 &       -1.12                 &           5\\
   IRAS\,F08572+3915&  264.0&  14.45&   2.97&   3.27&   3.29&    0.03/AGN                 &       -3.58                 &           4\\
          UGC\,04881&  178.0&   5.45&   3.22&   2.37&   2.99&     0.47/SF\tablenotemark{g}&       -0.63\tablenotemark{g}&           2\\
          UGC\,05101&  177.0&  10.23&   6.28&   1.99&   2.50&    0.13/AGN                 &       -0.78                 &           5\\
   IRAS\,F10173+0828&  224.0&   7.24&   2.58&   2.84&   3.10&     0.35/SF                 &       -1.20                 &           0\\
   IRAS\,F10565+2448&  197.0&  12.02&   5.84&   2.44&   3.04&     0.51/SF                 &       -0.75                 &           2\\
     MCG\,+07-23-019&  158.0&   4.17&   3.23&   2.37&   2.87&     0.64/SB                 &       -0.55                 &           2\\
   IRAS\,F11231+1456&  157.0&   4.37&   3.23&   2.65&   3.22&     0.60/SB                 &       -0.22                 &           1\\
           NGC\,3690&   50.8&   8.42&  47.24&   2.26&   2.71&    0.23/AGN\tablenotemark{g}&       -1.15\tablenotemark{g}&           3\\
   IRAS\,F12112+0305&  340.0&  22.91&   4.02&   2.66&   3.03&     0.30/SF                 &       -1.24                 &           4\\
          UGC\,08058&  192.0&  37.15&  14.22&   2.14&   2.30&    0.01/AGN                 &       -0.48                 &     \nodata\\
          UGC\,08387&  110.0&   5.37&   8.18&   2.32&   2.80&     0.62/SB                 &       -1.01                 &           4\\
        NGC\,5256\,S&  129.0&   3.09&   3.78&   1.90&   2.62&     0.44/SF                 &       -0.47                 &           3\\
          UGC\,08696&  173.0&  16.22&   9.75&   2.25&   2.78&    0.12/AGN                 &       -1.37                 &           5\\
   IRAS\,F14348-1447&  387.0&  24.55&   3.12&   2.37&   2.89&    0.25/AGN                 &       -1.36                 &           4\\
   IRAS\,F15163+4255&  183.0&   8.32&   4.34&   2.36&   2.91&     0.75/SB                 &        0.32                 &           4\\
   IRAS\,F15250+3608&  254.0&  12.02&   3.11&   2.76&   2.90&    0.03/AGN                 &       -2.69                 &           5\\
          UGC\,09913&   87.9&  19.05&  47.92&   2.59&   2.94&    0.17/AGN                 &       -2.26                 &           5\\
        NGC\,6286\,S&   85.7&   1.62&   5.08&   1.94&   2.63&     0.59/SB                 &       -0.40                 &     \nodata\\
           NGC\,7469&   70.8&   4.47&  13.31&   2.29&   2.85&    0.23/AGN                 &        0.06                 &           2\\
            IC\,5298&  119.0&   3.98&   4.46&   2.53&   3.16&    0.12/AGN                 &       -0.37                 &           0\\
            MRK\,331&   79.3&   3.16&   8.71&   2.52&   3.02&     0.63/SB                 &       -0.35                 &           1  
\enddata
\tablenotetext{a}{Distances and total IR ($8-1000\,\mu$m) luminosities are taken from \citet{lee09}.}
\tablenotetext{b}{IRAS-based far-infrared ($42.5-122.5\,\mu$m) fluxes taken from \citet{msc08}, where \( \left(\frac{F_{\rm FIR}}{\rm W\,m^{-2}}\right) = 1.26\times10^{-14}\left[\frac{2.58f_{\nu}(60\,\micron) + f_{\nu}(100\,\micron)}{\rm Jy}\right]\).}
\tablenotetext{c}{Logarithmic ratio of far-infrared-to-radio emission given in \citet{msc08}, where \(q_{\nu} = \log\left(\frac{F_{\rm FIR}}{3.75\times10^{12}~{\rm W\,m^{-2}}}\right) -\log\left(\frac{S_{\nu}}{\rm W\,m^{-2}\,Hz^{-1}}\right)\).}
\tablenotetext{d}{6.2\,$\mu$m PAH EQWs and silicate strengths ($s_{9.7\micron}$) are taken from \citep{ss13a}.  AGN and starburst (SB) dominated galaxies, based on having 6.2\,$\mu$m PAH EQWs below and above 0.27 and 0.54\,$\mu$m, respectively, are identified.  All systems with 6.2\,$\mu$m PAH EQWs above 0.27\,$\mu$m are considered to be star formation (SF) dominated.  }
\tablenotetext{e}{The merger stage as derived from the high resolution {\it HST} data (0 $=$ non-merger, 1 $=$ pre-merger, 2 $=$ ongoing merger with separable progenitor galaxies, 3 $=$ ongoing merger with progenitors sharing a common envelope, 4 $=$ ongoing merger with double nuclei plus tidal tail, 5 $=$ post-merger with single nucleus plus prominent tail, and 6 $=$ post-merger with single nucleus and a disturbed morphology), as described in \citet{haan11}.}
\tablenotetext{f}{Mid-infrared spectroscopy only available for the eastern component.}
\tablenotetext{g}{Luminosity-weighted average of mid-infrared measurements from individual galaxy nuclei.}
\end{deluxetable*}

%% file: tbl-2.tex
\begin{deluxetable*}{lcccccc}
\tablecaption{Radio Spectral Indices \label{tbl-2}}
\tabletypesize{\scriptsize}
\tablewidth{0pt}
\tablehead{
\colhead{Galaxy}  & \colhead{$<\nu_{low}>$\tablenotemark{a}} & \colhead{$\alpha_{low}$\tablenotemark{b}} &  \colhead{$<\nu_{mid}>$\tablenotemark{a}} & \colhead{$\alpha_{mid}$\tablenotemark{b}} &  \colhead{$<\nu_{high}>$\tablenotemark{a}} & \colhead{$\alpha_{high}$\tablenotemark{b}}\\
\colhead{}  & \colhead{(GHz)}  & \colhead{} & \colhead{(GHz)} & \colhead{} & \colhead{(GHz)} & \colhead{}
}
\startdata
             NGC\,34&    4.05~(2)&            0.70$\pm$0.031&    4.64~(3)&            0.74$\pm$0.029&   10.83~(3)&            0.87$\pm$0.014\\
            IC\,1623&    1.76~(3)&            0.90$\pm$0.057&    6.05~(3)&            0.85$\pm$0.024&   10.09~(3)&            0.95$\pm$0.087\\
       CGCG\,436-030&    2.18~(2)&            0.69$\pm$0.085&    4.63~(4)&            0.74$\pm$0.027&   14.58~(5)&            0.56$\pm$0.037\\
   IRAS\,F01364-1042&    2.44~(2)&            0.24$\pm$0.089&    5.07~(4)&            0.34$\pm$0.034&   12.88~(5)&            0.69$\pm$0.045\\
   IRAS\,F01417+1651&    2.22~(2)&            0.35$\pm$0.085&    4.13~(4)&            0.35$\pm$0.035&   15.74~(6)&            0.74$\pm$0.037\\
           UGC\,2369&    1.10~(3)&            0.33$\pm$0.079&    3.04~(2)&            0.73$\pm$0.058&   17.40~(2)&            0.93$\pm$0.117\\
   IRAS\,F03359+1523&    2.02~(3)&            0.50$\pm$0.073&    3.58~(3)&            0.34$\pm$0.051&   13.41~(3)&            0.68$\pm$0.073\\
           NGC\,1614&    1.16~(4)&            0.47$\pm$0.044&    7.12~(3)&            0.68$\pm$0.020&   10.10~(3)&            0.67$\pm$0.048\\
   IRAS\,F05189-2524&    2.75~(2)&            0.43$\pm$0.051&    3.73~(3)&            0.46$\pm$0.042&   11.19~(3)&            0.68$\pm$0.051\\
           NGC\,2623&    1.71~(3)&            0.33$\pm$0.084&    3.24~(3)&            0.53$\pm$0.055&   20.21~(3)&            0.71$\pm$0.075\\
   IRAS\,F08572+3915&    3.52~(3)&            0.02$\pm$0.081&    5.62~(4)&            0.02$\pm$0.057&   11.49~(5)&            0.26$\pm$0.047\\
          UGC\,04881&    1.10~(3)&            0.34$\pm$0.084&    4.03~(3)&            0.75$\pm$0.037&   12.18~(4)&            0.95$\pm$0.092\\
          UGC\,05101&    0.92~(5)&            0.51$\pm$0.031&    4.16~(4)&            0.70$\pm$0.037&   15.25~(6)&            0.93$\pm$0.047\\
   IRAS\,F10173+0828&    3.51~(3)&            0.36$\pm$0.063&    4.77~(3)&            0.38$\pm$0.069&    9.53~(4)&            0.48$\pm$0.065\\
   IRAS\,F10565+2448&    4.01~(4)&            0.73$\pm$0.022&    4.53~(3)&            0.77$\pm$0.030&    5.00~(2)&            0.81$\pm$0.179\\
     MCG\,+07-23-019&    2.50~(4)&            0.72$\pm$0.032&    4.28~(4)&            0.69$\pm$0.032&   12.79~(6)&            0.57$\pm$0.035\\
   IRAS\,F11231+1456&    1.17~(3)&            0.47$\pm$0.139&    4.09~(2)&            0.73$\pm$0.038&   11.84~(2)&            0.63$\pm$0.190\\
           NGC\,3690&    0.89~(5)&            0.61$\pm$0.027&    5.64~(4)&            0.59$\pm$0.034&   12.52~(5)&            0.69$\pm$0.041\\
   IRAS\,F12112+0305&    3.62~(4)&            0.37$\pm$0.028&    4.24~(3)&            0.38$\pm$0.029&    9.86~(4)&            0.73$\pm$0.025\\
          UGC\,08058&    3.39~(5)&            0.14$\pm$0.007&    5.49~(4)&            0.26$\pm$0.007&   10.07~(6)&            0.51$\pm$0.006\\
          UGC\,08387&    1.06~(4)&            0.50$\pm$0.019&    4.08~(4)&            0.60$\pm$0.035&   11.52~(4)&            0.67$\pm$0.063\\
        NGC\,5256\,S&    1.44~(5)&            0.72$\pm$0.028&    4.67~(3)&            0.91$\pm$0.030&   11.53~(4)&            0.66$\pm$0.059\\
          UGC\,08696&    0.89~(6)&            0.59$\pm$0.029&    4.21~(4)&            0.64$\pm$0.037&   13.38~(5)&            0.74$\pm$0.052\\
   IRAS\,F14348-1447&    1.29~(2)&           -0.22$\pm$0.244&    3.26~(2)&            0.67$\pm$0.055&   14.81~(2)&            0.89$\pm$0.147\\
   IRAS\,F15163+4255&    0.98~(3)&            0.72$\pm$0.053&    4.26~(3)&            0.68$\pm$0.034&   12.39~(3)&            0.87$\pm$0.075\\
   IRAS\,F15250+3608&    1.92~(2)&            0.09$\pm$0.190&    4.40~(4)&            0.14$\pm$0.040&   14.76~(6)&            0.71$\pm$0.051\\
          UGC\,09913&    1.51~(4)&            0.13$\pm$0.033&    4.06~(4)&            0.37$\pm$0.034&   14.62~(6)&            0.61$\pm$0.041\\
        NGC\,6286\,S&    1.05~(5)&            0.73$\pm$0.033&    4.35~(3)&            0.89$\pm$0.034&   10.81~(3)&            1.02$\pm$0.123\\
           NGC\,7469&    1.62~(3)&            0.63$\pm$0.042&    6.29~(3)&            0.71$\pm$0.018&   12.03~(3)&            1.06$\pm$0.031\\
            IC\,5298&    2.38~(2)&            0.74$\pm$0.087&    4.61~(3)&            0.80$\pm$0.035&   13.53~(3)&            0.80$\pm$0.056\\
            MRK\,331&    2.21~(2)&            0.67$\pm$0.085&    4.35~(3)&            0.65$\pm$0.030&   15.14~(3)&            0.78$\pm$0.045\\
\hline
             Medians&    1.76~(3)&            0.50$\pm$0.046&    4.35~(3)&            0.67$\pm$0.040&   12.39~(4)&            0.71$\pm$0.031
\enddata
\tablenotetext{a}{The average frequency, weighted by the signal-to-noise of each photometric data point, over which the radio spectral index was calculated.  The number in parentheses indicates how many frequencies were used in the calculation.}
\tablenotetext{b}{Radio spectral indices, where $S_{\nu} \propto \nu^{-\alpha}$, measured in the following frequency bins using data from \citet{msc10} and \citet{akl11b}; low: ($\nu < 5\,$GHz), mid: ($1\,{\rm GHz} < \nu < 10\,$GHz), and high: ($\nu > 4\,$GHz).}
\end{deluxetable*}